\documentclass[prd,twocolumn,
 superscriptaddress]{revtex4-2}
\usepackage{graphicx}% Include figure files
\usepackage{dcolumn}% Align table columns on decimal point
\usepackage{bm}% bold math
\usepackage[colorlinks=true, citecolor=blue, linkcolor=blue, urlcolor=blue]{hyperref}% add hypertext capabilities
\usepackage{xcolor}
\usepackage{amsmath, amssymb, amsfonts}
\usepackage{mathtools}
\usepackage{subfigure}
\usepackage{mathrsfs}
\usepackage{sidecap}
\usepackage{verbatim}

\definecolor{purple1}{rgb}{128,0,128}

\def\d{\mathrm{d}}

\begin{document}

\title{Analytical solutions for the quantum Brownian motion\\
of a particle during a quantum quench}

\author{Ygor \surname{de Oliveira Souza}}
\email{ygorosouza@unifei.edu.br}
\affiliation{Instituto de F\'{\i}sica e Qu\'{\i}mica, Universidade Federal de Itajub\'a, Itajub\'a, Minas Gerais 37500-903, Brasil}

\author{Caio C. \surname{Holanda Ribeiro}}
\email{caiocesarribeiro@alumni.usp.br}
\affiliation{International Center of Physics, Institute of Physics, University of Brasilia, 70297-400 Brasilia, Federal District, Brazil} 
\author{Vitorio A. \surname{De Lorenci}}
\email{delorenci@unifei.edu.br}
\affiliation{Instituto de F\'{\i}sica e Qu\'{\i}mica, Universidade Federal de Itajub\'a, Itajub\'a, Minas Gerais 37500-903, Brasil}

\date\today

\begin{abstract}
%A particle subjected to a fluctuating force due to its interaction to an external quantum system undergoes quantum Brownian motion. Here this phenomenon is considered in detail for a particle in a harmonic potential as it starts to interact with a reservoir modeled by a continuum of oscillators. By assuming a linear coupling between the particle and the reservoir, the system is canonically quantized and analytical expressions for the quantum correlations are found. These correlations are then applied to study the  system energy conservation and the particle Brownian motion is characterized from the perspective of the energy extracted from the reservoir.
A particle subjected to a fluctuating force originated from its interaction with an external quantum system undergoes quantum Brownian motion. This phenomenon is investigated in detail for the case of a particle confined by a harmonic potential and allowed to interact with a reservoir modeled as a continuum of oscillators, with particular emphasis on transient effects. Assuming a linear coupling between the particle and the reservoir, canonical quantization of the system is implemented, and analytical expressions for the quantum correlations are derived. These correlations are then applied to analyze energy conservation, and the particle Brownian motion is characterized from the perspective of the energy extracted from the reservoir.
\end{abstract}

\maketitle

\section{Introduction}
\label{sec2.2}

Since its first application to explain the black body radiation \cite{Rosenfeld}, quantum field theory has been responsible for some of the most interesting and counterintuitive predictions made by the scientific community, and it remains as a paradigm within the standard model of elementary fields. Among the impressive hallmarks of quantum theory applied to fields is the implementation of squeezed light in the LIGO detector \cite{Aasi2013}, that culminated in the first measurement of gravitational wave signals in 2016 \cite{PhysRevLett.116.061102}. Furthermore, quantum fields are expected to have a prominent role in black hole physics through the (semiclassical) mechanism of spontaneous Hawking radiation \cite{HAWKING1974}, that was also recently probed in an analogue black hole \cite{Jeff2019}.     

 Despite the numerous successful applications of quantum field theory, some problems of fundamental importance are still not fully addressed, specially in connection to gravity. For instance, energy conservation dictates that black holes loose mass due to Hawking radiation, and yet the question of how this energy extraction ends remains unanswered. In general, the problem of determining how quantum fluctuations affect their environment is convoluted and solutions valid only in certain regimes can be found. For instance, recent findings include quantum corrections due to a black hole formation \cite{PhysRevD.108.086012} and a solution to the backreaction problem in a Bose-Einstein condensate \cite{RibeiroPRA}.  

 Another phenomenon which is important to our analysis and that can be studied only approximately is the quantum Brownian motion of \cite{PhysRevD.70.065009}, in which charged test particles were shown to acquire velocity fluctuations due to their interaction with a quantum field. Specifically, in the regime of non-relativistic dynamics, a particle of mass $m$ and charge $q$ is released at $t=0$ and at a distance $d$ from a plane perfect mirror. If the particle position does not vary appreciably, then its velocity undergoes fluctuations analytically given by 
\begin{align*}
    \langle v_{\bot}^2\rangle&=\frac{q^2}{\pi^2m^2}\frac{t}{32d^3}\ln\left(\frac{2d+t}{2d-t}\right)^2,\\
    \langle v_{\|}^2\rangle&=\frac{q^2}{8\pi^2m^2}\left[\frac{t}{8d^3}\ln\left(\frac{2d+t}{2d-t}\right)^2-\frac{t^2}{d^2(t^2-4d^2)}\right],
\end{align*}
where $v_\bot$ and $v_\|$ are the velocity components orthogonal and parallel to the mirror, respectively, and units are such that $\varepsilon_0=\hbar=c=1$. 

Apart from the divergences in the above formulas, whose sources are well-understood (see, for instance, \cite{PhysRevD.107.076007} and references therein), two features deserve mentioning. First, the assumption on the particle movement being negligible forbids the analysis of the late-time (ballistic) regime. Indeed, electromagnetic vacuum fluctuations lead to a diffusive stochastic dynamics \cite{PhysRevD.90.027702}, showing that larger particle displacements become relevant as time passes. Second, and more important, $\langle v^2_{\|}\rangle<0$ for $t>2d$. This kind of phenomenon, where a classically positive definite quantity becomes negative upon quantization, is not uncommon in quantum field theory, e.g., the electromagnetic energy density in the Casimir effect, and it is usually called a {\it subvacuum} effect. However, for a particle's velocity, one always expects positive dispersions, as these are linked to uncertainties in measurements. For the particular case of the quantum Brownian motion of \cite{PhysRevD.70.065009}, it is conjectured that, once the test particle quantum features are taken into account, the overall velocity dispersion becomes a positive number and the effect of the mirror is to diminish the magnitude of this number. 

In this work we discuss how the quantum Brownian motion of a particle is modified as it starts to interact with a reservoir. Specifically, we consider a Lagrangian model for a particle under a harmonic potential and a reservoir modeled by a continuum of oscillators, using for the latter the dielectric model of \cite{barnett1992} as motivation. By performing the canonical quantization of the system, analytical expressions for the quantum correlations are found. Therefore, the late-time regime and the positivity of the particle kinetic energy can be fully addressed within our model, as no approximation is needed in determining the quantum dynamics.

It should be stressed that the damped harmonic oscillator is one of the most studied systems of quantum optics and it serves as a paradigm for the description of various systems in nature, and, in particular, for the study of the quantum Brownian motion \cite{PhysRevB.110.085114,PhysRevA.108.012210,PhysRevA.105.012209,PhysRevLett.116.120402,PhysRevLett.108.170404,PhysRevE.84.041139,CACHEFFO20102198,DIAS200973,PhysRevE.77.011112,PhysRevA.66.042118,PhysRevD.64.105020,PhysRevA.61.022107,PhysRevD.49.6612}. Included in the vast literature on the subject, we cite, for instance, \cite{PhysRevD.45.2843,FLEMING20111207}, where a master equation for the density operator was developed and analytical solutions were found, and \cite{Ford1988-2}, where the method of Langevin equation is developed. For a recent account on the subject, see \cite{PhysRevResearch.4.033151} and references therein. Our approach follows a distinct path, to the extent that the total system is quantized canonically. Although this procedure is in general involved and cannot lead to analytical results, it is necessary when one needs to have full control of the quantum correlations. In our particular case, we are interested in determining how the conservation of energy occurs in the system, which can become convoluted or even impossible to address if reduced density operator and Langevin methods are adopted.  

The work is organized as follows. In section \ref{themodel} the Lagrangian model is introduced alongside the details of the system initial condition. Section \ref{secenergyconservation} discusses energy conservation in the system, and establishes a notion for the particle energy that can be used as a probe of the quantum Brownian motion. In section \ref{seccanonicalquantization} the system is canonically quantized, and in section \ref{secexactcorr} the two-point function is presented. Section \ref{secqbm} then discussed the quantum Brownian motion in detail, and the work finishes with some remarks in section \ref{secfinal}. Throughout the analysis units are such that $\hbar=1$.

\section{The model Lagrangian}
\label{themodel}

We consider a single non-relativistic particle of mass $m$ in a harmonic potential of characteristic frequency $\omega_0$ with Lagrangian
\begin{equation}
L_{\rm p}=\frac{m}{2}\dot{x}^2-\frac{m\omega_0^2}{2}x^2,
\end{equation}
such that $x=x(t)$ is the particle position, and hereafter a dot over a quantity means time-derivative. We assume that this particle interacts with a reservoir modeled by a continuum of oscillators labeled by $\nu$, $\nu>0$, with Lagrangian 
\begin{equation}
L_{\rm R}=\frac{\mu}{2}\int_0^\infty\d\nu\left(\dot{R}^2-\nu^2R^2\right).
\end{equation}

The reservoir variables are such that $R=R(t,\nu)$. We take the possibly time-dependent interaction between the test oscillator and the reservoir to be mediated by
\begin{equation}
L_{\rm int}=x\int_0^\infty\d\nu\beta\dot{R},\label{interactionlagrangian}
\end{equation}
where the coupling constant, $\beta=\beta(t,\nu)$, is assumed to be different from zero for all reservoir frequencies $\nu$. Also, $\beta$ is taken to satisfy $\beta(t,-\nu)=\beta(t,\nu)$. 
This type of coupling is well-known and it is the essence, for instance, of the Hopfield dielectric model \cite{barnett1992}. The Lagrangian $L=L_{\rm p}+L_{\rm R}+L_{\rm int}$ gives rise to the equations of motion
\begin{align}
&\ddot{x}+\omega_0^2x=\frac{1}{m}\int_0^\infty\d\nu\beta\dot{R},\label{eqphi}\\
&\ddot{R}+\nu^2R=-\frac{1}{\mu}\frac{\d\ }{\d t}(\beta x).\label{eqr}
\end{align}
We see that if $\beta\neq0$, Eq.~\eqref{eqphi} describes a harmonic oscillator driven by the external force $F=\int_0^\infty\d\nu\beta\dot{R}$, which in turn depends on the oscillator velocity $v=\dot{x}$ by means of Eq.~\eqref{eqr}. Therefore, the coupling given by Eq.~\eqref{interactionlagrangian} models some sort of viscous medium for the oscillator under study. Note also that this coupling is distinct from the coupling of \cite{FEYNMAN1963118,CALDEIRA1983587}, but is included in the generalized model of \cite{CALDEIRA1983374}.

Notice that when the reservoir oscillators are quantized, the force $F$ undergoes quantum fluctuations, giving rise to a sort of Quantum Brownian motion \cite{Ford1988}. In order to gain a deeper insight on how this occurs, let us consider the (weak coupling) limit where $\beta\rightarrow0$, for which the rhs of Eq.~\eqref{eqr} approaches zero. In physical terms, this corresponds to the case where the oscillator under study is  a test particle that cannot interfere with the reservoir in an appreciable manner. In this limit, the momentum canonically conjugated to $R$ is $Q=\mu\dot{R}+\beta x\approx \mu\dot{R}$
%
%\begin{align}
%p&=\frac{\delta L}{\delta \dot{x}}=m\dot{x},\\
%Q&=\frac{\delta L}{\delta %\dot{R}}=M\dot{R}+\beta x,
%\end{align}
%
and let us assume first that only the reservoir oscillators are quantized, i.e., an operator-valued distribution for $R(t,\nu)$ subjected to $[R(t,\nu),Q(t,\nu')]=i\delta(\nu-\nu')$ is known. Then the force $F$ is also an operator and we find that Eq.~\eqref{eqphi} becomes a Langevin equation for $x$ leading to $\langle(\Delta v)^2\rangle\neq0$ by means of the fluctuations of $F$. Here, the expectation value is taken with respect to the reservoir quantum state. This is precisely the regime considered in the quantum Brownian motion of \cite{PhysRevD.70.065009}, where the force $F$ was sourced by Casimir stresses on a charged test particle.

In our model, though, we do not assume the weak coupling regime and we also consider the particle's own ``quantumness.'' Thus uncertainties in measuring the particle velocity exist independently of the particle's interaction with the reservoir. Therefore, from the knowledge of $\langle(\Delta v)^2\rangle$ as a function of time, one can, in principle, determine how the interaction with the reservoir eventually affects the particle's Brownian motion.

The considerations above are general and hold for arbitrary time-dependencies of the coupling constant. Nevertheless, in our work we are interested in transient effects when the particle's interaction with the reservoir starts at a given time, say, $t=0$. Specifically, we consider the case for which $\beta=0$ for $t<0$, henceforth called the non-interacting period, and $\beta$ assumes a time-independent value after $t>0$, the interacting regime. This assumption has the advantage of leading to analytical solutions for the quantum correlations.

\section{Energy conservation}
\label{secenergyconservation}

The major advantage of adopting microscopic models, which are notoriously convoluted, over Langevin methods is the level of control over the conservation laws. Following \cite{DeLorenci2019}, we take the particle mechanical and kinetic energy as probes of the quantum fluctuations on the system. The system Hamiltonian obtained from $L$ is:
\begin{align}
    H=&\frac{p^2}{2m}+\frac{m\omega_{\rm e}^2}{2}x^2-\frac{x}{\mu}\int_0^\infty\d\nu\beta Q\nonumber\\
    &+\frac{1}{2}\int_0^{\infty}\d\nu\left(\frac{Q^2}{\mu}+\mu\nu^2R^2\right),\label{energymomentum}
\end{align}
where $p=m\dot{x}$ is the particle momentum, 
%$H=H_{\rm p}+H_{\rm r}+H_{\rm int}$, with
%
%\begin{align}
%    H_{\rm p}&=\frac{p^2}{2m}+\frac{m\omega_{\rm e}^2}{2}x^2,\\
%    H_{\rm r}&=\frac{1}{2}\int_0^{\infty}\d\nu\left(\frac{q_\nu^2}{m_r}+m_r\nu^2r_{\nu}^2\right),\\
%    H_{\rm int}&=-\frac{x}{m_r}\int_0^\infty\d\nu\beta(\nu)q_\nu,
%\end{align}
%
and the time-dependent ``effective'' frequency $\omega_{\rm e}$ is defined by
\begin{equation}
    \omega_{\rm e}^2=\omega_0^2+\frac{1}{m \mu}\int_0^\infty\d\nu\beta^2.\label{effectivefrequency}
\end{equation}
We note that because the system is interacting there is some degree of freedom in interpreting the particle's energy, the effective frequency, the interaction energy, and the reservoir energy, because only the full Hamiltonian, $H$, is unambiguously defined. The quantity $\omega_{\rm e}$ only coincides with the oscillator effective frequency in certain cases, where a careful limit of the system parameters should be observed. We cite \cite{dodonov2023} for a recent study of a particle in a time-dependent harmonic potential.

For our purposes, and because we work in the Heisenberg picture, it is instructive to express the Hamiltonian in terms of velocities rather than momenta as in Eq.~\eqref{energymomentum}, which will allow for the identification of how the particle mechanical energy changes as the interaction is turned on. We find that $H=H_{\rm p}+H_{\rm R}$, where 
\begin{align}
    H_{\rm p}&=\frac{m}{2}\dot{x}^2+\frac{m\omega_0^2}{2}x^2,\label{energy}\\
    H_{\rm R}&=\frac{\mu}{2}\int_0^\infty\d\nu\left(\dot{R}^2+\nu^2R^2\right).
\end{align}
We note that $H_{\rm p}$ coincides with the particle's energy before the interaction is turned on, at $t=0$. Furthermore, the equations of motion imply that
\begin{equation}
\frac{\d H}{\d t}=-x\int_0^\infty\d\nu \dot{\beta}\dot{R}, \label{energyconservationlaw}   
\end{equation}
and thus $\d H/\d t=0$ for $t\neq0$, when $\beta$ is time-independent. Accordingly, changes in $H_{\rm p}$ are followed by changes in $H_{\rm R}$ for $t\neq0$ ensuring that $H$ remains constant.
We will adopt the observable $H_{\rm p}$ as a measure of how the particle energy changes as it enters the interacting regime.

In what follows, the quantization of this theory is worked out for all possible choice of parameters in order to determine the fluctuations of $x$ and $\dot{x}$, the probes of the environment-induced Brownian motion on the particle. We are interested in studying how $\langle H_{\rm p}\rangle$ changes as the interactions are turned on. We also discuss changes in $\langle T\rangle$, where $T=mv^2/2$ is the particle kinetic energy. It should be stressed that the identification of the particle's energy is one of the main strengths of the microscopic model over Langevin equation methods, for which in general it is not possible to write down the system Hamiltonian and it is not always clear how to treat time-dependent scenarios. This is similar to what occurs in discussing energy in effective models for polarization and magnetization in electrodynamics \cite{wald2022}.

\section{Canonical Quantization}
\label{seccanonicalquantization}

We work in the Heisenberg picture, for which quantization can be obtained from the expansion of the quantities $x$ and $R$ in a complete set of mode functions followed by imposition of the canonical commutation relations on the Fourier coefficients. Therefore, we start by solving for the most general solution of the equations of motion.

We define the two-component ``field'' $\Psi$, such that $\Psi_1=\Psi_1(t)=x(t)$ and $\Psi_2=\Psi_2(t,\nu)=R(t,\nu)$. We note that $\nu$ plays the same coordinate role as $t$. With this definition, Eqs.~\eqref{eqphi} and \eqref{eqr} combine into a single field equation for $\Psi$, which in turn implies that the sesquilinear form
\begin{align}
\langle\Psi,\Psi^{'}\rangle=i\bigg[m&(\Psi_1^*\partial_t\Psi_1^{'}-\Psi_1^{'}\partial_t\Psi_1^*)\nonumber\\
&+\mu\int_0^\infty\d\nu\left(\Psi_2^*\partial_t\Psi_2^{'}-\Psi_2^{'}\partial_t\Psi_2^*\right)\nonumber\\
&-\int_0^\infty\d\nu\beta(\Psi_1^{*}\Psi_2^{'}-\Psi_1^{'}\Psi_2^{*})\bigg],\label{scalarproduct}
\end{align}
is time-independent for {\it any two solutions} of the field equation. We can use this scalar product to find a complete set of positive norm field modes $\{\Psi_\omega\}_\omega$, i.e., $\langle\Psi_\omega,\Psi_{\omega'}\rangle=\delta_{\omega\omega'}$ such that
\begin{equation}
\Psi=\sum_\omega\left[a_\omega\Psi_\omega+a^*_\omega\Psi^*_\omega\right],
\end{equation}
and $a_\omega=\langle\Psi_\omega,\Psi\rangle$. We note that $\omega$ here is a generic index that might assume continuous and/or discrete values. Also, it follows from Eq.~\eqref{scalarproduct} that if $\Psi_\omega$ is a positive norm solution, then $\Psi_\omega^*$ has negative norm, and both positive and negative norm modes are necessary to span the whole space of solutions. Finally, the Fourier coefficients $a_\omega$ are time-independent and uniquely determined once $\Psi$, $\partial_t\Psi$ are given in some initial instant of time, i.e., we have a well-defined Cauchy problem. 

As customary in any field theory, different sets of field modes lead to physically distinct physical vacua. In our model, a privileged choice can be made in the non-interacting regime, for which the theory vacuum corresponds to {\it all} the oscillators in their fundamental states. Specifically, for $t<0$, we look for field modes indexed by their positive frequency $\omega$ such that $\Psi_{\omega}\propto\exp(-i\omega t)$. These functions comprise a complete set of positive norm field modes.

Indeed, the first obvious field mode is given $\omega=\omega_0$, with $\Psi_{\omega_0,2}=0$, and
\begin{equation}
\Psi_{\omega_0,1}(t)=\frac{1}{\sqrt{2\omega_0 m}}e^{-i\omega_0 t},
\end{equation}
which is already normalized: $\langle\Psi_{\omega_0},\Psi_{\omega_0}\rangle=1$. The second family of mode functions correspond to the case where only one reservoir oscillator is excited. If we denote by $\Phi_{\omega}$ such functions, then, for each $\omega>0$, we find $\Phi_{\omega,1}=0$, and 
\begin{equation}
\Phi_{\omega,2}(t,\nu)=\frac{\delta(\nu-\omega)}{\sqrt{2\nu \mu}}e^{-i\nu t}.
\end{equation}
Therefore, for $t<0$, we find the general expression
\begin{equation}
\Psi=a_{\omega_0}\Psi_{\omega_0}+a^{*}_{\omega_0}\Psi_{\omega_0}^{*}+\int_0^\infty\d\omega(b_{\omega}\Phi_{\omega}+b^{*}_{\omega}\Phi_{\omega}^{*}).\label{psiexp}
\end{equation}
In particular, because $x=\Psi_1$, we find for $t<0$ that
\begin{equation}
x(t)=\frac{1}{\sqrt{2\omega_0 m}}(a_{\omega_0}e^{-i\omega_0 t}+a^{*}_{\omega_0}e^{i\omega_0 t}).
\end{equation}

We observe that each mode function is a solution of the field equation. Thus, if the evolution of $\Psi_{\omega_0}$ and $\Phi_{\omega}$ is known, Eq.~\eqref{psiexp} furnishes $x$ at all times. In order to find the evolution of $\Psi_{\omega_0}$, we note that if $\{\Gamma_{\omega}\}_{\omega}$ is a complete set of positive norm field modes in the interaction period, then, for $t>0$, we can write 
\begin{equation}
\Psi_{\omega_0}(t)=\int_0^\infty\d\omega[c_{\omega}\Gamma_{\omega}(t)+d^{*}_{\omega}\Gamma^{*}_{\omega}(t)],\label{fourierpsi}
\end{equation} 
where $c_{\omega}=\langle\Gamma_{\omega},\Psi_{\omega_0}\rangle|_{t\rightarrow0^{+}}$, $d^{*}_{\omega}=-\langle\Gamma^{*}_{\omega},\Psi_{\omega_0}\rangle|_{t\rightarrow0^{+}}$. Note that as the interacting period starts, the second entry of $\Psi_{\omega_0}$, that vanishes in the non-interacting period, can be non-zero. Similarly, we find that, for $t>0$,
\begin{equation}
\Phi_{\omega}(t)=\int_0^\infty\d\omega'[c_{\omega,\omega'}\Gamma_{\omega'}(t)+d^{*}_{\omega,\omega'}\Gamma^{*}_{\omega'}(t)],\label{fourierphi}
\end{equation}
and $c_{\omega,\omega'}=\langle\Gamma_{\omega'},\Phi_{\omega}\rangle|_{t\rightarrow0^{+}}$, $d^{*}_{\omega,\omega'}=-\langle\Gamma^{*}_{\omega'},\Phi_{\omega}\rangle|_{t\rightarrow0^{+}}$.

The complete set $\{\Gamma_\omega\}$ can be obtained as follows. By assuming a time-dependence in the form $\Gamma_\omega(t)=\exp(-i\omega t)\Gamma_\omega^{0}$ with $\omega>0$, we find that
\begin{align}
&(\omega_0^2-\omega^2)\Gamma^{0}_{\omega,1}+\frac{i\omega}{m}\int_0^\infty\d\nu\beta\Gamma^{0}_{\omega,2}=0,\label{eqpart1}\\
&(\nu^2-\omega^2)\Gamma^{0}_{\omega,2}-\frac{i\omega}{\mu}\beta\Gamma^{0}_{\omega,1}=0.\label{eqpart2}
\end{align}
Now, the general solution of Eq.~\eqref{eqpart2} is
\begin{equation}
\Gamma^{0}_{\omega,2}(\nu)=\frac{i\omega}{\nu^2-\omega^2}\frac{\beta(\nu)}{\mu}\Gamma^{0}_{\omega,1}+A_{\omega}\delta(\nu-\omega),
\end{equation}
where $A_\omega$ is an arbitrary constant. Thus, Eq.~\eqref{eqpart1} implies that
\begin{equation}
A_{\omega}=\frac{im}{\omega\beta(\omega)}\zeta_r(\omega)\Gamma^{0}_{\omega,1},
\end{equation}
with $\zeta_r(\omega)=\mbox{Re}[\zeta(\omega)]$, and we have defined the complex function
\begin{equation}
\zeta(\omega)=\omega_0^2-\omega^2-\frac{\omega}{2m\mu}\int_{-\infty}^{\infty}\d\nu\frac{\beta^2(\nu)}{\nu-\omega+i\epsilon}\label{zetadef}.
\end{equation}
Here, $\epsilon>0$ is a small parameter to the taken to zero at the end of the calculations. Clearly, when $\omega>0$, 
\begin{equation}
\zeta_i(\omega)=\mbox{Im}[\zeta(\omega)]=\pi\frac{\omega\beta^2(\omega)}{2m\mu}>0.
\end{equation}
The quantity $\Gamma^{0}_{\omega,1}$ is a normalization constant, which can be determined through the scalar product. In fact, we find that
\begin{equation}
\langle\Gamma_\omega,\Gamma_{\omega'}\rangle=\left|\Gamma^{0}_{\omega,1}\right|^2\frac{m\pi|\zeta(\omega)|^2}{\zeta_i(\omega)}\delta(\omega-\omega'),
\end{equation}
and thus
\begin{equation}
\Gamma^{0}_{\omega,1}=\frac{1}{\sqrt{m\pi}}\frac{\sqrt{\zeta_i(\omega)}}{\zeta(\omega)}
\end{equation}
finishes the construction of the complete set of positive norm mode functions $\{\Gamma_\omega\}_{\omega}$.

Therefore, the various Fourier components appearing in Eqs.~\eqref{fourierpsi}, \eqref{fourierphi} are given by 
\begin{align}
c_\omega=&\sqrt{\frac{m}{2\omega_0}}\left.\left(\omega_0\Gamma^{*}_{\omega,1}-i\partial_t\Gamma^{*}_{\omega,1}\right)\right|_{t=0}%=\frac{1}{\sqrt{2\omega_0\pi}}(\omega+\omega_0)\frac{\sqrt{\zeta_i(\omega)}}{\zeta^{*}(\omega)}
,\\
d^*_\omega=&-\sqrt{\frac{m}{2\omega_0}}\left.\left(\omega_0\Gamma_{\omega,1}-i\partial_t\Gamma_{\omega,1}\right)\right|_{t=0}%=\frac{1}{\sqrt{2\omega_0\pi}}(\omega-\omega_0)\frac{\sqrt{\zeta_i(\omega)}}{\zeta(\omega)}
,\\
c_{\omega,\omega'}=&\left.\sqrt{\frac{\mu}{2\omega}}\left[(\omega-i\partial_t)\Gamma^{*}_{\omega',2}(t,\omega)-\frac{i\beta(\omega)}{\mu}\Gamma^{*}_{\omega',1}\right]\right|_{t=0},\\
d^{*}_{\omega,\omega'}=&\left.\sqrt{\frac{\mu}{2\omega}}\left[(i\partial_t-\omega)\Gamma_{\omega',2}(t,\omega)+\frac{i\beta(\omega)}{\mu}\Gamma_{\omega',1}\right]\right|_{t=0}.
\end{align}

Finally, the quantization of the system is concluded by promoting the c-numbers $a_{\omega_0}$, $b_{\omega}$ of the expansion \eqref{psiexp} to operators subjected to the commutation relations $[a_{\omega_0},b_{\omega}]=0,[b_{\omega},b_{\omega'}]=0$, and
\begin{align}
&[a_{\omega_0},a^\dagger_{\omega_0}]=1,\\
&[b_{\omega},b^\dagger_{\omega'}]=\delta(\omega-\omega').
\end{align} 
Accordingly, the system vacuum state, $|0\rangle$, defined by $a_{\omega_0}|0\rangle=b_{\omega}|0\rangle=0$ for all $\omega>0$, corresponds to the case where all the oscillators are in their fundamental states before the interaction is turned on.

We note that by plugging the mode expansions \eqref{fourierpsi}, \eqref{fourierphi} back into Eq.~\eqref{psiexp} we find that, for $t>0$,
\begin{equation}
\Psi(t)=\int_0^\infty\d\omega\left[\gamma_\omega\Gamma_\omega(t)+\gamma^\dagger_\omega\Gamma^*_\omega(t)\right],\label{quasiexp}
\end{equation}
where the operators $\gamma_\omega$ are related to the non-interacting creation operators via the Bogoliubov transformation \cite{Birrell_Davies_1982}
\begin{equation}
\gamma_{\omega}=c_{\omega}a_{\omega_0}+d_{\omega}a^{\dagger}_{\omega_0}+\int_0^\infty\d\nu\left(c_{\nu,\omega}b_\nu+d_{\nu,\omega}b^\dagger_\nu\right).\label{bogo}
\end{equation}
A lengthy computation then reveals that the operators $\gamma_\omega$ satisfy
\begin{align}
\left[\gamma_\omega,\gamma_{\omega'}\right]&=0,\\
\left[\gamma_\omega,\gamma^\dagger_{\omega'}\right]&=\delta(\omega-\omega').
\end{align}
The commutation relations above are of great importance to our analysis. Indeed, if the system under study were quantized already in the interacting period, one would obtain the expansion \eqref{quasiexp} with the corresponding (instantaneous) vacuum state, $|0\rangle_{\rm qp}$, of the theory defined as $\gamma_\omega|0\rangle_{\rm qp}=0$ for all $\omega$. This choice of quasiparticle vacuum gives rise to the well-known model explored in \cite{Ford1988,Ford1988-2}. In our case, however, $|0\rangle_{\rm qp}\neq|0\rangle$, and thus Eq.~\eqref{bogo} is the Bogoliubov transformation relating the two quantum field representations. We discuss the relation between the two vacua in the next section.    

We finish this section with a remark regarding the canonical commutation relation for the Heisenberg operator [see Eq.~\eqref{quasiexp}]
\begin{equation}
    x(t)=\frac{1}{\sqrt{m\pi}}\int_0^\infty\d\omega\sqrt{\zeta_i(\omega)}\left[\gamma_\omega\frac{e^{-i\omega t}}{\zeta(\omega)}+H.c.\right].
\end{equation}
Thus it follows that
\begin{equation}
    [x(t),p(t)]=-\frac{1}{\pi}\int_{-\infty}^{\infty}\d\omega\frac{\omega}{\zeta(\omega)},\label{ccr}
\end{equation}
where we used the definition $p=m\dot{x}$. We note that the function $\zeta(\omega)$ defined in Eq.~\eqref{zetadef} is an analytic function in the lower half complex $\omega$ plane. Furthermore, $\zeta(\omega)\neq 0$ for all $\omega$ satisfying $\mbox{Im}(\omega)\leq0$. This can be seen as follows. If we write $\omega=\omega_r-i\omega_i$, with $\omega_i>0$, we find that $\mbox{Im}[\zeta(\omega)]=0$ only if $\omega_r=0$, whereas
\begin{equation}
    \zeta(-i\omega_i)=\omega_0^2+\omega_i^2\left(1+\frac{1}{m\mu}\int_0^\infty\d\nu\frac{\beta^2(\nu)}{\nu^2+\omega_i^2}\right),
\end{equation}
is always positive for $\omega_i>0$. We conclude from this that $1/\zeta(\omega)$ is also analytic in the lower half-plane. This can be used to evaluate the integral in Eq.~\eqref{ccr} by closing the integration contour in the lower half complex plane to obtain
\begin{equation}
    [x(t),p(t)]=-\frac{i}{\pi}\lim_{r\rightarrow\infty}\int_\pi^{2\pi}\d\theta\frac{r^2e^{2i\theta}}{\zeta(re^{i\theta})}=i,
\end{equation}
where we used the property $\zeta(\omega)\rightarrow-\omega^2$ for $|\omega|\rightarrow\infty$, which holds as long as $\beta(\nu)\rightarrow 0$ for $\nu\rightarrow\infty$.

\section{Two-point correlation function for an exactly solvable case}
\label{secexactcorr}

We now turn our attention to the two-point function $\langle x(t)x(t')\rangle$, which can be used to calculate the oscillator quantities of interest here. %Henceforth we restrict our attention to the particular case where $1/\zeta$ has only {\it isolated simple poles} in the upper half-plane. We let $\sigma=\{\Omega\in\mathbb{C}:\zeta(\Omega)=0\}$ be the set of poles of $1/\zeta$ and we denote the residue of $1/\zeta(\omega)$ at $\omega=\Omega\in\sigma$ as $R_\Omega$. Accordingly, we show in the Appendix \ref{appendixA} that
%
%\begin{equation}
%\frac{1}{\zeta(\omega)}=%\frac{-1}{2i\pi}\int_{-\infty}^{\infty}\d\omega'\frac{1}{\zeta(\omega')(\omega'-\omega+i\epsilon)}=
%\sum_{\Omega\in\sigma}\frac{R_{\Omega}}{\omega-\Omega},
%\end{equation} 
%
%i.e., $1/\zeta$ is completely characterized by its residues, from which we obtain the correlation 
%
By using the analytical properties of $1/\zeta(\omega)$, one can show that the correlation function assumes the form
\begin{equation}
\langle x(t)x(t')\rangle=\langle x(t)x(t')\rangle_{\rm tr}+\langle x(t)x(t')\rangle_{\rm qp},\label{quantumcorrfull}
\end{equation}
valid for $t,t'\geq0$, where
\begin{align}
\langle x(t)x(t')\rangle_{\rm tr}=&\frac{1}{2\omega_0m}\bigg\{[(\omega_0+i\partial_t)A(t)][(\omega_0-i\partial_{t'})A^*(t')]\nonumber\\
&\hspace{0cm}+\frac{2\omega_0}{\pi}\int_0^\infty\d\omega\zeta_i(\omega)\bigg[\frac{e^{-i\omega t}}{\zeta^*(\omega)}B^*_{\omega}(t')\nonumber\\
&\hspace{0cm}+B_{\omega}(t)\frac{e^{i\omega t'}}{\zeta(\omega)}+B_{\omega}(t)B^*_{\omega}(t')\bigg]\bigg\},
\end{align}
%\begin{widetext}
%\begin{align}
%\langle x(t)x(t')\rangle_{\rm tr}=\frac{1}{2\omega_0m}\sum_{\Omega,\Omega'}R_{\Omega}R_{\Omega'}\Bigg[&(\Omega-\omega_0)(\Omega'+\omega_0)e^{i\Omega t+i\Omega' t'}\nonumber\\
%&+\frac{2\omega_0}{\pi}\int_0^\infty\d\omega\frac{\zeta_i(\omega)}{(\omega-\Omega')(\omega+\Omega)}\left(e^{i\Omega t+i\omega t'}+e^{-i\omega t+i\Omega' t'}-e^{i\Omega t+i\Omega' t'}\right)\Bigg],
%\end{align}
%\end{widetext} 
%
is the transient correlation and
\begin{equation}
\langle x(t)x(t')\rangle_{\rm qp}=\frac{1}{m\pi}\int_0^\infty\d\omega\frac{\zeta_i(\omega)}{|\zeta(\omega)|^2}e^{-i\omega\Delta t},\label{wightmanqp}
\end{equation}
is the two-point function with respect to the instantaneous quasiparticle vacuum state. Here we defined the auxiliary functions
\begin{align}
    A(t)&=\frac{1}{\pi}\int_{-\infty}^\infty\d\omega\frac{\sin\omega t}{\zeta(\omega)},\\
    B_{\omega}(t)&=\frac{1}{2\pi i}\int_{-\infty}^\infty\d\omega'\frac{e^{i\omega't}}{\zeta(\omega')}\frac{1}{\omega'+\omega+i\epsilon}.
\end{align}

In general, depending on the coupling constant $\beta(\nu)$, the analytical properties of $1/\zeta(\omega)$ in the upper half complex plane can be convoluted. For instance, we note that for $\beta(\nu)^2\propto\Theta(\nu_0^2-\nu^2)$, $1/\zeta(\omega)$ will have logarithmic dependence on $\omega$. An interesting scenario that can be exactly integrated and produces a meromorphic $1/\zeta(\omega)$ occurs when the coupling constant $\beta$ is the combination of two Lorentzians:
\begin{equation}
    \frac{\beta^2(\omega_0\eta)}{\omega_0}=\frac{\sigma^2m\mu}{\pi}\left[\frac{\eta_0}{(\eta-\eta_r)^2+\eta_0^2}+\frac{\eta_0}{(\eta+\eta_r)^2+\eta_0^2}\right],\label{lorentiziancoupling}
\end{equation}
which represents a coupling with maximum intensity at $\omega_0\eta=\omega_0\eta_r$. Here $\sigma$ is a dimensionless constant that measures the magnitude of the interaction. Also, $\eta_0\rightarrow0$ implies
\begin{equation}
    \frac{\beta^2(\omega_0\eta)}{\omega_0}\rightarrow\sigma^2m\mu\left[\delta(\eta-\eta_r)+\delta(\eta+\eta_r)\right].\label{resonance}
\end{equation}
The corresponding oscillator effective squared frequency is [cf.~Eq.~\eqref{effectivefrequency}]
\begin{equation}
    \omega_{\rm e}^2=\omega_0^2(1+\sigma^2),
\end{equation}
and 
\begin{equation}
    \frac{\zeta(\omega_0\eta)}{\omega_0^2}=1-\eta^2+\eta\sigma^2\frac{\eta-i\eta_0}{(\eta-i\eta_0)^2-\eta_r^2}.\label{zetaint}
\end{equation}
%
%where $\eta_0=\nu_0/\omega_0$, $\eta_r=\nu_r/\omega_0$. 

%
The advantage of the functional form of Eq.~\eqref{zetaint} is that $1/\zeta(\omega)$ is a meromorphic function with exactly four simple poles in the upper half plane, determined by the three independent parameters $\sigma, \eta_r,\eta_0$. Furthermore, these poles are the zeros of $\zeta(\omega_0\eta)$, which are determined by a degree four polynomial equation in $\eta$. We let $\eta_i$, $i=1,2,3,4$, be these complex roots, and $\mathcal{R}_{\eta_i}$ be the corresponding residue of $\omega_0^2/\zeta(\omega_0\eta)$ at $\eta_i$. Also, the reflection property $\zeta(\omega)^*=\zeta(-\omega)$ implies that $-\eta_i^*$ is also a root with residue $\mathcal{R}_{-\eta_i^*}=-\mathcal{R}_{\eta_i}^*$. These properties hold for any meromorphic $1/\zeta$ with simple poles. For such functions,
\begin{align}
    \frac{\omega_0^2}{\zeta(\omega_0\eta)}&=\sum_{i}\frac{\mathcal{R}_{\eta_i}}{\eta-\eta_i},
\end{align}
which is a consequence of the Residue Theorem and can be used to write the stationary two-point function \eqref{wightmanqp} as
\begin{align}
    \langle x(t)&x(t')\rangle_{\rm qp}=\frac{-i}{\omega_0m\pi}\sum_j\mathcal{R}_{\eta_j}\Big[\sin(\eta_j\omega_0\Delta t)\mbox{Si}(\eta_j\omega_0\Delta t)\nonumber\\
    &+\cos(\eta_j\omega_0\Delta t)\mbox{Ci}(\eta_j\omega_0\Delta t)-\frac{\pi}{2}\sin(\eta_j\omega_0\Delta t)\Big],\label{corrqp}
\end{align}
where $\mbox{Si}$ and $\mbox{Ci}$ are the sine and cosine integral functions, defined as
\begin{align}
\mbox{Si}(z) &= \int_0^z \frac{\sin(t)}{t} \d t = \frac{\pi}{2} - \int_z^\infty \frac{\sin(t)}{t}\d t,\\
\mbox{Ci}(z)&=-\int_z^\infty \frac{\cos(t)}{t} \d t.
\end{align}
Also, we set $\Delta t=t-t'-i\epsilon$, where $\epsilon>0$ should be taken to zero later on, at the end of the calculations.
%
%\begin{equation}
%F(\Omega,\Delta t)=\int_0^\infty\d\omega \frac{\omega}{\omega^2-\Omega^2}e^{-i\omega \Delta t}.
%\end{equation}
%
%This function can be integrated exactly in terms of special functions.

Finally, the transient part of the two-point function is exactly given by
\begin{align}
&\langle x(t)x(t')\rangle_{\rm tr}=\frac{1}{2\omega_0m}\sum_{k,j}\mathcal{R}^*_{\eta_k}\mathcal{R}_{\eta_j}\times\nonumber\\
\Bigg\{&\left[(\eta_k^*+1)(\eta_j+1)-\frac{\sigma^2}{\pi}F_{kj}(0)\right]e^{-i\eta_k^* \omega_0t+i\eta_j \omega_0t'}\nonumber\\
&+\frac{\sigma^2}{\pi}\left[e^{i\eta_j\omega_0 t'}F_{kj}(\omega_0 t)+e^{-i\eta_k^*\omega_0 t}F^*_{jk}(\omega_0 t')\right]\Bigg\},\label{corrtr}
\end{align}
where the various auxiliary functions are defined as
\begin{align}
    g(\alpha)=&e^{i \alpha}\left[\frac{i\pi}{2}+\mbox{Ci}(\alpha)-i\mbox{Si}(\alpha)\right],\\
    G_{kj}(c,\alpha)=&\frac{cg(c\alpha)+\eta_k^*g(-\eta_k^*\alpha)}{(\eta_j-\eta_k^*)(\eta_k^*+c)}+(\eta_j\leftrightarrow\eta_{k}^*),\\
    %&+\frac{cg(c\alpha)+\eta_jg(-\eta_j\alpha)}{(\eta_k^*-\eta_j)(\eta_j+c)},\\
    F_{kj}(\alpha)=&\frac{i}{2}[G_{kj}(-\eta_r-i\eta_0,\alpha)+G_{kj}(\eta_r-i\eta_0,\alpha)]\nonumber\\
    %&\hspace{-0.1cm}-G_{kj}(-\eta_r+i\eta_0,\alpha)-G_{kj}(\eta_r+i\eta_0,\alpha)].
    &-(\eta_0\leftrightarrow -\eta_0).
\end{align}
Equations \eqref{corrqp} and \eqref{corrtr} represent one of the main results in our work. We stress that although the damped harmonic oscillator is one of the most studied systems in quantum optics, the above equations represent an exact solution for the Wightman function in such systems, that can be used to unveil various quantum features of interest.

\section{Quantum Brownian Motion}
\label{secqbm}

We are now able to discuss in detail the quantum Brownian motion. For definiteness, we assume in this work that before the quench the particle and the reservoir oscillators are in their fundamental state, at zero temperature. Accordingly, $\langle x\rangle\equiv0$ and $\langle v\rangle\equiv0$ throughout the system evolution. We are interested in determine how the oscillator changes as it starts to interact with the environment. Thus, if $\mathcal{E}_0$ is the particle energy at $t<0$, then $\Delta \mathcal{E}=\langle H_{\rm p}\rangle-\mathcal{E}_0$ probes how much the particle energy changed. Similarly, if $\mathcal{T}_0$ is the particle kinetic energy at $t<0$, $\Delta \mathcal{T}=\langle T\rangle-\mathcal{T}_0$ is the corresponding change due to the reservoir.

\subsection{Late-time regime}

The parameters $\eta_0,\eta_r$ and $\sigma$ determine how the particle achieves the late-time regime, when transients effects already took place. According to Eq.~\eqref{energy}, the change in particle energy, on average, after the system reaches its equilibrium is $\Delta\mathcal{E}_{\rm asy}=\lim_{t\rightarrow\infty}\Delta\mathcal{E}$, which reads, with the aid of the correlation function \eqref{corrqp},
\begin{align}
\Delta\mathcal{E}_{\rm asy}=&\frac{m}{2}\lim_{t'\rightarrow t}(\partial_t\partial_{t'}+\omega_0^2)\langle x(t)x(t')\rangle_{\rm qp}-\frac{\omega_0}{2}\nonumber\\
=&\frac{\omega_0}{2}\bigg[\frac{-i}{\pi}\sum_j\mathcal{R}_{\eta_j}(1+\eta_j^2)\ln(-i\eta_j)-1\bigg],
\end{align}
whereas the change in the particle kinetic energy, $\Delta\mathcal{T}_{\rm asy}=\lim_{t\rightarrow\infty}\Delta\mathcal{T}$, reads
\begin{equation}
   \Delta\mathcal{T}_{\rm asy}= \frac{\omega_0}{2}\bigg[\frac{-i}{\pi}\sum_j\mathcal{R}_{\eta_j}\eta_j^2\ln(-i\eta_j)-\frac{1}{2}\bigg].
\end{equation}

\begin{figure}[h!]
\center
\includegraphics[width=0.48\textwidth]{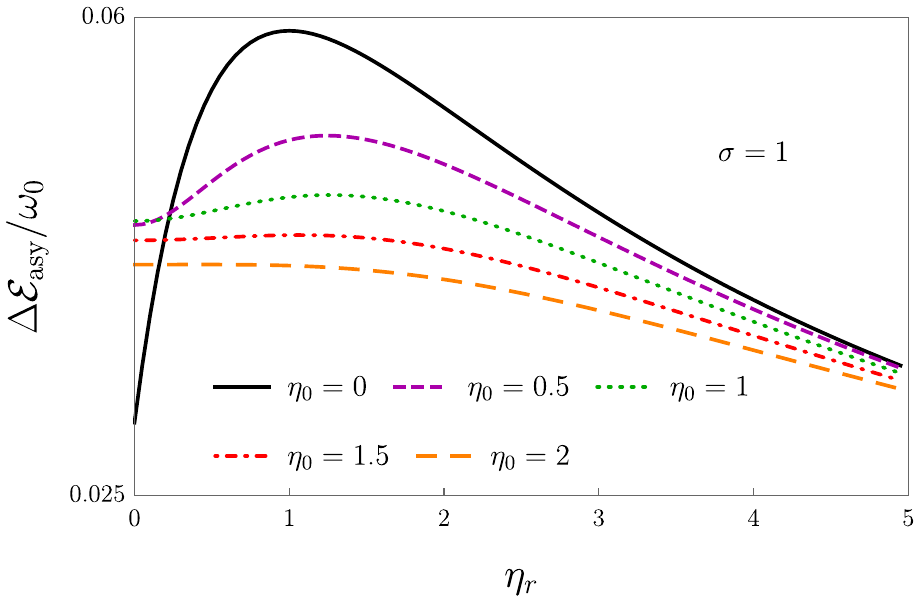}
\caption{Late-time behavior of the particle energy. Here it is assumed that $\sigma=1$. Notice that the particle energy gain is maximum near $(\eta_0,\eta_r)\sim(0,1)$, which corresponds to a resonance with the reservoir oscillator of frequency $\nu=\omega_0$. Also, note that the energy exchange is minimum for $\eta_0=\eta_r\sim0$, the regime in which the particle exchange energy only with the low energy reservoir oscillators.}
\label{fig1}
\end{figure}

\begin{figure}[h!]
\center
\includegraphics[width=0.48\textwidth]{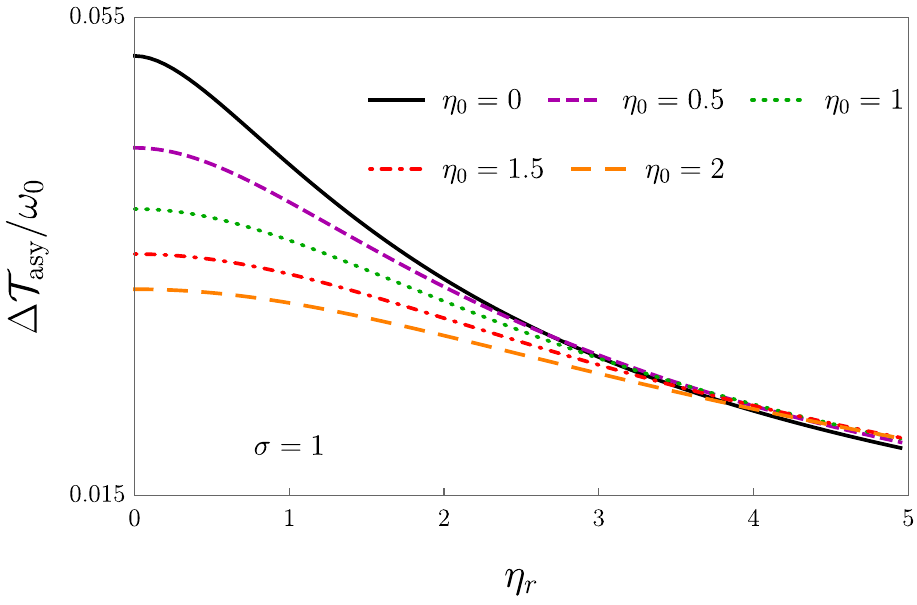}
\caption{Late-time behavior of the particle kinetic energy, for $\sigma=1$. Notice that, differently from the total energy, the particle kinetic energy is maximum at around $\eta_0=\eta_r\sim0$, and it is insensitive to the resonance at $\eta_r=1$.}
\label{fig2}
\end{figure}

Figures \ref{fig1} and \ref{fig2} depict $\Delta\mathcal{E}_{\rm asy}$ and $\Delta\mathcal{T}_{\rm asy}$, respectively, as functions of $\eta_0$ and $\eta_r$, for a fixed $\sigma=1$. Figure \ref{fig1} shows that the particle energy always increases due to the interaction with the reservoir and this effect is more pronounced for $(\eta_0,\eta_r)\sim(0,1)$, i.e., near the resonance [cf.~Eq.~\eqref{resonance}]. 
Also, although it is fairly intuitive that the particle energy gain is maximum around the resonance, Fig.~\ref{fig1} also reveals that the energy gain is minimum (for the parameter window shown in the figure) near $\eta_0=\eta_r\sim0$. We attribute this to the fact that in this regime the particle is allowed to exchange energy mainly with the low energy reservoir oscillators.

Figure \ref{fig2} depicts the late-time behavior of the particle kinetic energy exchange. Similarly to the total particle energy, the kinetic energy always increases due to the interaction with the reservoir. Moreover, interesting features are observed in sharp distinction to the total energy: the particle kinetic energy is insensitive to the resonance at $\eta_r=1$ and it is maximum near $\eta_0=\eta_r\sim0$, where the particle interacts mostly with low energy reservoir oscillators.

\subsection{Transient regime}

We now consider the transient regime after the quantum quench. We start with some remarks about the transient time duration as function of the parameters $\sigma$, $\eta_0$ and $\eta_r$. Inspection of Eq.~\eqref{corrtr} shows that the relaxation times are determined by the imaginary parts of the roots $\eta_j$, and these roots are solution of the degree four polynomial equation
\begin{equation}
    (1-\eta^2)[(\eta-i\eta_0)^2-\eta_r^2]+\sigma^2\eta(\eta-i\eta_0)=0.\label{poleq}
\end{equation}
Moreover, the contribution of a given root $\eta_j$ to the transient correlation is modulated by the residue $\mathcal{R}_{\eta_j}$, and thus an interesting interplay between these quantities occur. For $\sigma\rightarrow0$ (weak coupling regime), the solutions of Eq.~\eqref{poleq} read
\begin{align}
    \eta&=\pm 1+\frac{\sigma^2}{2}\frac{\pm1-i\eta_0}{(\pm1-i\eta_0)^2-\eta_r^2}+\mathcal{O}(\sigma^4),\\
    \eta&=\pm\eta_r+i\eta_0+\frac{\sigma^2}{2}\frac{\pm\eta_r+i\eta_0}{(\pm\eta_r+i\eta_0)^2-1}+\mathcal{O}(\sigma^4),
\end{align}
whereas for $\sigma\rightarrow\infty$ (strong coupling), we find
\begin{align}
    \eta&=\pm \sigma+\frac{i\eta_0}{2}+\mathcal{O}(\sigma^{-1}),\\
    \eta&=\frac{i}{\sigma^2}\frac{\eta_0^2+\eta_r^2}{\eta_0}+\mathcal{O}(\sigma^{-3}),\\
    \eta&=i\eta_0-\frac{i}{\sigma^2}\frac{\eta_r^2(\eta_0^2+1)}{\eta_0}+\mathcal{O}(\sigma^{-3}).
\end{align}
Thus, these formulas can be used to determine the relaxation times in each of the asymptotic regimes. For any $\sigma>0$, the minimum among the imaginary parts of the $\eta_j$, when multiplied by $\omega_0$, determines the system relaxation time.
\begin{figure}[h!]
\center
\includegraphics[width=\linewidth]{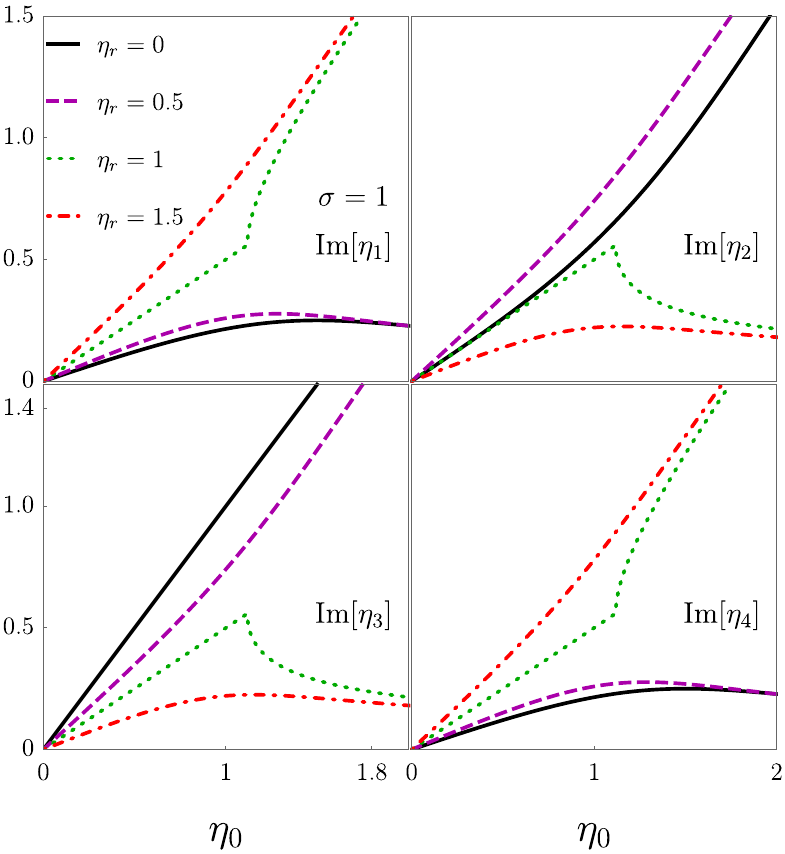}
\caption{Imaginary parts of the roots of Eq.~\eqref{poleq} for $\sigma=1$. The labels of the roots are not relevant for our analysis. Note that as $\eta_0\rightarrow0$, the imaginary parts vanish, meaning that the system takes longer to reach equilibrium for smaller $\eta_0$.}
\label{fig3}
\end{figure}
Figure \ref{fig3} depicts the imaginary parts of the roots for $\sigma=1$. 

\begin{figure}[h!]
\center
\includegraphics[width=0.48\textwidth]{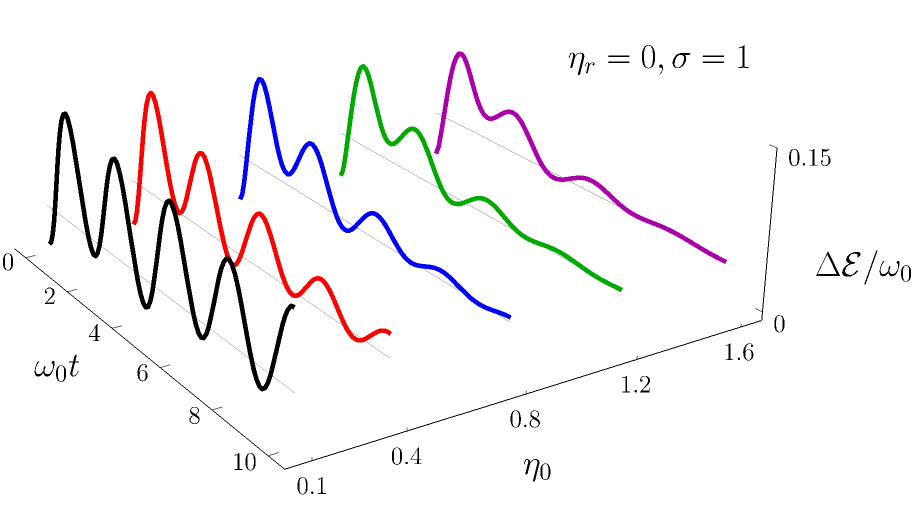}
\includegraphics[width=0.48\textwidth]{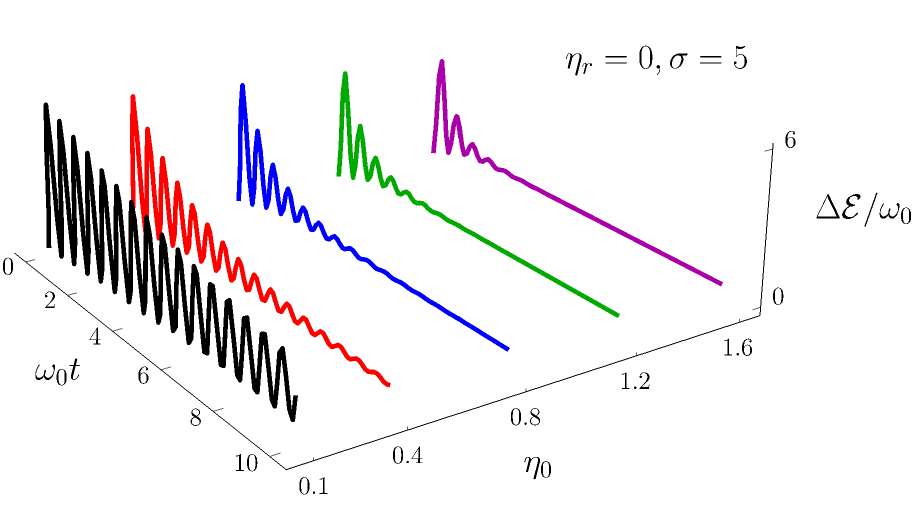}
\includegraphics[width=0.48\textwidth]{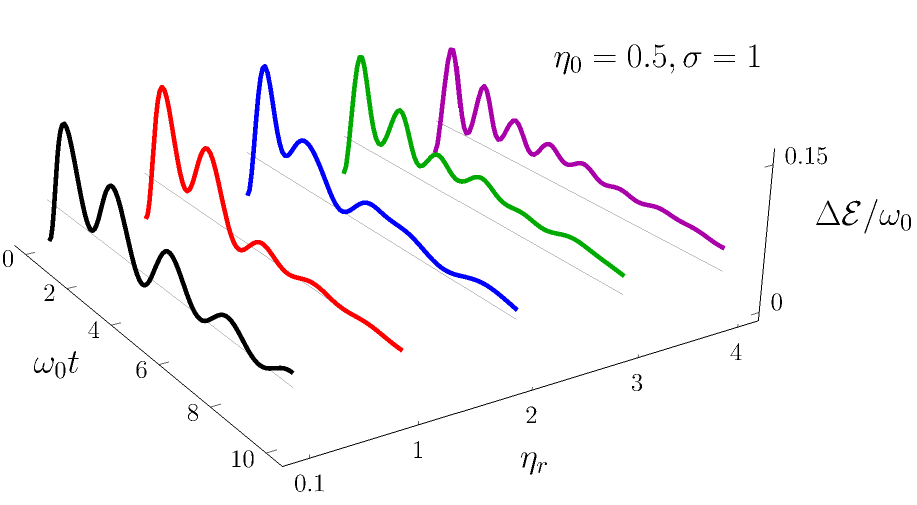}
\caption{Particle energy variation as function of time. Top panel: 
 Particle energy for $\eta_r=0$ and $\sigma=1$. Notice that the relaxation time decreases with $\eta_0$ (for fixed $\sigma$). Middle panel: Particle energy for $\eta_r=0$ and $\sigma=5$. Notice that the amplitude and the frequency of the oscillations increase with $\sigma$. Bottom panel: Particle energy for $\eta_0=0.5$ and $\sigma=1$. 
 The resonance parameter $\eta_r$ has a non-trivial effect on the particle energy.}
\label{fig4}
\end{figure}

Note that, in general, the relaxation time increases as $\eta_0$ goes to zero. An example of this behavior is depicted in Fig.~\ref{fig4} upper panel, for $\eta_r=0$, $\sigma=1$. One can see from the figure that as the interaction is turned on, the particle energy acquires a damped oscillatory behavior as function of time, with a frequency determined by $\sigma$. Figure \ref{fig4} middle panel depicts the particle energy for the same $\eta_0$ and $\eta_r$ parameters and for the stronger coupling of $\sigma=5$. Notice that both the frequency of the oscillations and their magnitude increase as $\sigma$ grows.  
We also present in Fig.~\ref{fig4} bottom panel the particle energy as function of time for several values of $\eta_r$ and for $\eta_0=0.5$, $\sigma=1$, from which one can see that $\eta_r$, which determines the frequency of the reservoir oscillators that are in resonance with the particle, also affects the oscillatory pattern. 

\begin{figure}[h!]
\center
\includegraphics[width=0.48\textwidth]{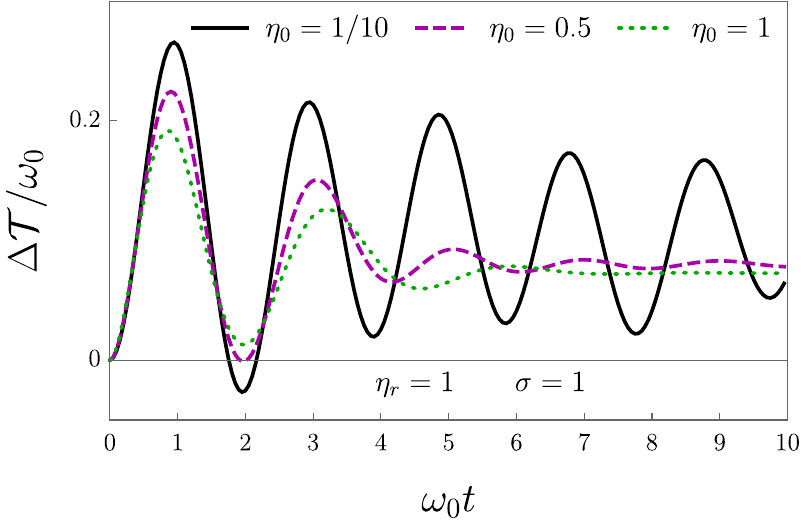}
\caption{Particle kinetic energy variation as function of time, for $\eta_r=\sigma=1$ and several values of $\eta_0$. Notice that depending on the system parameters, the particle can actually loose some of its initial kinetic energy, in analogy to the subvacuum effect of \cite{PhysRevD.70.065009}.}
\label{fig6}
\end{figure}
As a final application of the analytical correlations, we present in Fig.~\ref{fig6} how the particle kinetic energy changes during the transient regime. Recall that in the semi-classical quantum Brownian motion discussed in \cite{PhysRevD.70.065009}, the particle kinetic energy, which is assumed to be initially zero, can actually become negative due to the quantum fluctuations of an external field. In our model, which contain an analytical solution to the quantum Brownian motion, the analogue effect is the diminishing of the particle's initial (positive) kinetic energy. 

Figure \ref{fig6} shows a situation where this occurs, for $\eta_{0}=0.1, \eta_r=\sigma=1$. For these parameters, the particle, after interacting with the reservoir for some period of time, actually looses part of its kinetic energy. We note that in this system this is a true quantum effect, whose origin can be traced back to the phenomenon of quantum squeezing \cite{doi:10.1126/science.aac5138}. Indeed, note that the initial kinetic energy decreases if and only if $\langle p^2\rangle$ decreases, at the expense of increasing $\langle x^2\rangle$ to ensure the validity of Heisenberg's uncertainty relation.

We finish this section with a remark regarding the conservation of energy in this system. Figure \ref{fig4} shows that the particle, for the parameters considered in the plots, always gain energy. There are two possible origins to this energy gain, namely, the particle can extract energy from the reservoir and it can gain energy from the external agent that turns the interaction on at $t=0$. Here, because the microscopic model is known, it is possible to pinpoint exactly the source of the particle energy gain. Indeed, it follows from Eq.~\eqref{energyconservationlaw} that the particle and the reservoir energies, $\langle H_{\rm p}\rangle$ and $\langle H_{\rm R}\rangle$, respectively, are related through
\begin{align}
    \frac{\d\langle H_{\rm p}\rangle}{\d t}+\frac{\d\langle H_{\rm R}\rangle}{\d t}=-\int_0^\infty\d\nu \dot{\beta}\langle x\dot{R}\rangle.
\end{align}
Therefore, because the rhs of the above equation is zero for $t\neq0$ and $H_{\rm p}$ is a continuous function of time, it follows that the energy gained by the particle comes exclusively from its interaction with the reservoir oscillators.

\section{Final Remarks}
\label{secfinal}

In this work we studied how a reservoir modifies the quantum Brownian motion of a particle. By starting from a Lagrangian model for the total system, we were able to quantize the system canonically and obtain analytical solutions for the quantum correlations. The major results in our work are the analytical correlations and their application to characterize the quantum Brownian motion of a particle. 

A couple of important remarks are in order. We note that the linear coupling between the particle and the reservoir is an important assumption in order to find analytical solutions. In general, for non-linear couplings the system cannot be quantized analytically and methods to find approximate solutions are necessary. For instance, for the quantum Brownian motion of \cite{PhysRevD.70.065009} the particle interacts non-linearly with the electric field, and the assumption of negligible particle displacement was implemented in order to find approximate solutions.

Also, we note that the reservoir model here implemented serves as an analogue model to the electric field of \cite{PhysRevD.70.065009}, to the extent that it can simulate important features like the subvacuum effect and gives information about the late-time regime. However, this analogy is only qualitative, for the system of \cite{PhysRevD.70.065009} is a Casimir-like system imprinting velocity fluctuations onto a charged particle, for which, due to its complexity, still remains not thoroughly studied.

We conclude this work with a remark regarding the generality of the two-point function \eqref{quantumcorrfull}. The analysis of the quantum Brownian motion was perform for a particular type of interaction given by the two Lorentzians of Eq.~\eqref{lorentiziancoupling}, that leads to analytical solutions. Nevertheless, the correlations \eqref{quantumcorrfull} can be used to study quantum properties of the particle for all types of couplings as long as $\beta\rightarrow 0$ when $\nu\rightarrow 0$, which is a consistency condition for the function $\zeta$ given in Eq.~\eqref{zetadef} to be well-defined for all $\omega>0$.

\section*{Acknowledgements}
C.C.H.R. would like to thank the {\it Funda\c{c}\~ao de Apoio \`a Pesquisa do Distrito
Federal} under grant 00193-00002051/2023-14. V.A.D.L. is supported in part by {\it Conselho Nacional de Desenvolvimento Cient\'{\i}fico e Tecnol\'ogico} under grant  302492/2022-4.

%where $\mathcal{L}$, $\mathcal{C}$, and $\mathcal{G}$ are defined as
%
%\begin{align}
%&\mathcal{L}=\lim_{\Delta x\rightarrow0^+}\frac{L_n}{\Delta x},\ \ \mathcal{C}=\lim_{\Delta x\rightarrow0^+}\frac{C_{n+1}}{\Delta x},\\
%&\mathcal{G}=\lim_{\Delta x\rightarrow0^+}\frac{L_{nj}}{\Delta x^2}.
%\end{align} 
%

% 
%\begin{figure}[h!]
%\center
%\includegraphics[width=0.45\textwidth]{fig2.pdf}
%\caption{Dispersion relation. Each line of constant $\ell\omega/v$ intercepts the curves at the real wave vector solutions of the dispersion relation. Positive slopes (positive %group velocity) correspond to rightwards propagation whereas negative slopes, to leftwards propagation. For $\theta=0$ the system is in the vicinity of a dynamical instability.}
%\label{fig2}
%\end{figure}
%

%\appendix

%\section{A curious relationship for $1/\zeta(\omega)$}
%\label{appendixA}

%Notice that
%
%\begin{align}
%    \frac{1}{\zeta(\omega)}&=-\frac{1}{2i\pi}\int^{\infty}_{-\infty}\frac{\d\nu}{\zeta(\nu)(\nu-\omega+i\epsilon)}\nonumber\\
%    &=\sum_{\Omega\in\sigma}\frac{R_{\Omega}}{\omega-\Omega}.
%\end{align}

\bibliography{refs}

\end{document}